\documentclass{amia}
\usepackage{graphicx}
\usepackage[labelfont=bf]{caption}
\usepackage{color}
\usepackage[colorinlistoftodos,prependcaption]{todonotes}
\usepackage{regexpatch}
\usepackage{graphicx}
\usepackage{tabularx}
\usepackage{hyperref}
\usepackage{cleveref}
\usepackage{natbib}
\usepackage[utf8]{inputenc}
\usepackage{booktabs}
\usepackage{multirow}
\usepackage[normalem]{ulem}
\useunder{\uline}{\ul}{}
\usepackage{caption}
\usepackage{subcaption}

\usepackage[font=itshape,begintext=``,endtext='']{quoting}
\quotingsetup{vskip=-2pt}

\let\OLDthebibliography\thebibliography
\renewcommand\thebibliography[1]{
  \OLDthebibliography{#1}
  \setlength{\parskip}{0pt}
  \setlength{\itemsep}{0pt plus 0.3ex}
}

\usepackage{listings}

\usepackage{titlesec}
\titlespacing*{\section}{0pt}{1.1\baselineskip}{\baselineskip}

\begin{document}

\title{BlockIoT: Blockchain-based Health Data Integration using IoT Devices}

\author{Manan Shukla$^{1}$, Jianjing Lin, PhD$^{1}$, Oshani Seneviratne, PhD$^{1}$}

\institutes{
    $^1$Rensselaer Polytechnic Institute, Troy, NY, USA\\
}
\maketitle

\noindent{\bf Abstract}


The development and adoption of Electronic Health Records (EHR) and health monitoring Internet of Things (IoT) Devices have enabled digitization of patient records and has also substantially transformed the healthcare delivery system in aspects such as remote patient monitoring, healthcare decision making, and medical research. However, data tends to be fragmented among health infrastructures, and prevents interoperability of medical data at the point of care. 
In order to address this gap, we introduce BlockIoT that uses blockchain technology to transfer previously inaccessible and centralized data from medical devices to EHR systems, which provides greater insight to providers who can, in turn, provide better outcomes for patients. This notion of interoperability of medical device data is possible through an Application Programming Interface (API), which serves as a versatile endpoint for all incoming medical device data, a distributed file system that ensures data resilience, and knowledge templates that analyze, identify, and represent medical device data to providers. Our participatory design survey on BlockIoT demonstrates that BlockIoT is a suitable system to supplement physicians’ clinical practice and increases efficiency in most healthcare specialties, including cardiology, pulmonology, endocrinology, and primary care.

\section*{Introduction}
In the last several decades, the explosion of information technology has profoundly changed how information is stored, exchanged, managed, and analyzed in medicine. For instance, the invention of EHRs has enabled digitization of patient records and has substantially transformed the healthcare delivery system. At the same time, the emergence of IoT health devices has also been increasingly applied in remote patient monitoring, healthcare decision making, and medical research by offering a variety of physiological data (such as heart rate, EKG, blood sugar levels) of the patient~\cite{medical-device-market}.
The advancement of these technologies has created the potential to improve the quality and efficiency in the healthcare industry. However, clinical data is still primarily stored and managed in a fragmented manner, which creates friction in information exchange at the point of care and also hinders large-scale health-data research empowered by technology such as artificial intelligence/machine learning (AI/ML). This paper focuses on the integration between EHRs and health-monitoring IoT devices in a decentralized environment, in the hope of offering a potential solution to this challenge. 

As of 2020, almost every patient's health information has been stored in some EHR system, and an increasing number of patients carry medical IoT devices, especially those with chronic diseases. However, we believe that the true value of medical devices has not been realized, given the minimal application of physiological data in clinical practices. Such a limited application could arise from various reasons, including conflicting incentives from different parties, imperfect regulations, technology barriers, etc. Our paper aims to provide a potential solution in the technical aspect of this issue. The foremost important issue is lack of \emph{interoperability} between medical devices and clinical data infrastructures, such as the EHR system, where providers mainly store and retrieve patient health information. As a result, medical device data is not accessible by most providers when it could have been helpful to reach a proper clinical decision.
Moreover, data security has always been a significant concern. For instance, storing data in a centralized system has been criticized to expose sensitive data to cyber-attacks. According to the Healthcare Data Breach report, millions of patient medical records are compromised every month~\cite{healthcare-data-breach-report}.
Concerns on data security and patient privacy make the transmission of health information even more cumbersome and costly, which further impedes the improvement of healthcare decision making and the advancement of medical research. 

To assuage these concerns, we propose an innovative information transfer system combining a blockchain-based technology and an API that enables interoperability between IoT medical devices and EHR systems, titled \emph{BlockIoT}. 
%
%
First, BlockIoT serves as a method that allows medical devices to quickly and easily interface with the EHRs.
It removes a significant amount of overhead (in the form of labor/development costs necessary to modify existing medical device protocols) necessary to transfer data between different parties that use individual proprietary systems to manage data. The proposed system will allow individual medical device data to be securely transferred to providers without changing any firmware on medical devices or forcing companies to modify existing protocols when creating medical devices. As a result, it simplifies the required infrastructures for data transmission and reduces the number of middlemen involved, which could potentially improve cost efficiency in this process.
Another advantage of BlockIoT is the ability to present providers with essential data points by communicating and analyzing medical device data. For instance, providers or insurance companies can use algorithms and visual diagrams to quickly understand medical data and determine whether the patient’s health is within normal limits or requires medical attention. This allows for quicker healthcare interventions, which prevents unnecessary and costly visits (such as to the emergency room) and allows patients’ symptoms to be resolved at a much faster pace than in the current healthcare setting.
Notably, the proposed system can be combined with emerging technology such as AI/ML to improve the quality and efficency of healthcare delivery. For example, it can be connected to an automatic alert system facilitated by AI technology to generate messages to providers about unexpected abnormal symptoms or to remind patients to comply with medication. Studies have shown that simple reminders or alerts in the form of text messages or emails can change patients' behavior and improve their health as a result \citep{roter1998effectiveness, neff2009periodic, vann2018patient}. However, due to constraints in time and resources, such interventions from providers are often not readily available and thus, patients have to undertake most of the responsibility to stay compliant and follow the proper treatment regimen, which is particularly challenging for elderly patients with multiple complicated comorbidities.

We assess the proposed system by conducting structured interviews with healthcare practitioners from various backgrounds. During the interview, we first introduced BlockIoT and then presented a mockup of a sample EHR system integrated with BlockIoT. We also provided a History of Present Illness (HPI) description of a sample patient, who suffers from multiple complications and carries different types of medical devices. Then we asked providers opinions on how to treat the sample patient in the presence and absence of the device data. We also asked for suggestions on what additional information is desired and how the presentation of information can be improved.
The majority of respondents viewed medical device data as enabling a more complete and objective picture of patients' health. Moreover, the system can improve providers' understanding of patients' health and facilitate communication with patients. Adopting such a system can streamline the healthcare delivery workflow as long as the system is user-friendly and information can be presented clearly and succinctly. Our clinician evaluators, in general, acknowledge the value of the system, especially if the associated devices are also easy to use for patients.

\section*{Related Work}

There have been several attempts at bridging blockchain, EHR, and wearable health IoT devices. However, our literature review indicated that the success of these systems had been limited. This section illustrates some of the comparable related works and how BlockIoT transcends the capabilities offered by them.

A system for remotely monitoring patients in a Health Insurance Portability and Accountability Act (HIPAA) compliant manner is introduced in \cite{Rights_OCR_2008, griggs2018healthcare}. However, the system does not focus on specific methods or protocols a device can use to communicate with the blockchain,
forces device manufacturers to change device specifications to conform to the system's requirements, no method is described that enables medical devices to integrate into the system, or in regards to representing this data to a physician.
A data transfer system between a physician and patient through the use of a blockchain system for the transfer of labs and medical charts (rather than medical device data) is described in \cite{ramani2018secure}. However, unlike in our system, this transfer of data is only one-sided and does not allow any analysis of the data present, which subjects the blockchain integration to be simply another form of a database. 
The integration of IoT devices to mobile applications is discussed in \cite{dharmendra2017iot}. The framework manages medical devices but is very shallow in terms of its data transfer and storage. Primarily, representing data in a mobile application that is only available to a patient severely limits the potential use of the data itself. Secondly, because the system is centralized rather than decentralized, a significant risk is present in which any breach of that system can result in the release of significant amounts of medical data. Finally, there is no method present at all that can analyze or encrypt incoming data. 
An EHR system that can connect various IoT devices is introduced in \cite{vuppalapati2016role}. However, the method it uses to achieve this has severe limitations. Data is primarily stored in a centralized rather than a decentralized system, which results in a significant risk in which any breach of that system can result in the release of significant amounts of medical data. Secondly, while data can be analyzed, the system can be integrated with only specific medical devices and support a minimal scope of incoming data. 

On improving EHR security, a system that enhances the security of current blockchains without the use of any specific keys is described in \cite{nagasubramanian2020securing}.
However, they do not provide practical methods to input or output data from the system. Instead, it forces a physician to manually input data into the system itself.
Using blockchain to secure medical device data and prevent any malware or breaches from occurring is introduced in \cite{paliokas2019blockchain}. However, enforcement of this security comes with significant limitations as well. Primarily, no method is present to communicate medical data to another party, preventing physicians access to patient data. Secondly, similar to other papers above, the system forces manufacturers to create an entirely new device to satisfy the system's needs, which prevents any commercial device to use the system.

On data sharing beyond the clinical settings, a secure method to transfer data between a researcher, patient, and physician through the use of blockchain is investigated in \cite{shen2019medchain}. However, the paper does have some limitations. Primarily, no method or protocol exists for medical devices to send data in real-time to the system. Instead, the system leaves this data to the medical device company or the patients to do on their own. This process significantly discourages the transfer of medical device data from the patient to the relevant physician, and the data itself is unable to be analyzed in real-time, which prevents any life-threatening alerts from reaching the patient.

\section*{Methodology}

The BlockIoT system is developed with four main functional goals. (1) Receiving real-time raw data from various commercial but standards-compliant medical devices; (2) converting raw data-points into \textit{well-defined and structured information}, which are stored in a decentralized system; (3) analyzing raw data-points through smart contracts for potential patient or physician alerts; and (4) releasing this data to various EHR systems to represent medical data in a concise and easy-to-read fashion. These four aspects serve as a basis for the implemented work to increase patient data interoperability between medical devices and EHR systems. We first introduce the system's architecture and the various parties involved and detail specific operations that highlight interoperability.

\begin{figure}[!htbp]
\centering
\includegraphics[width=\columnwidth]{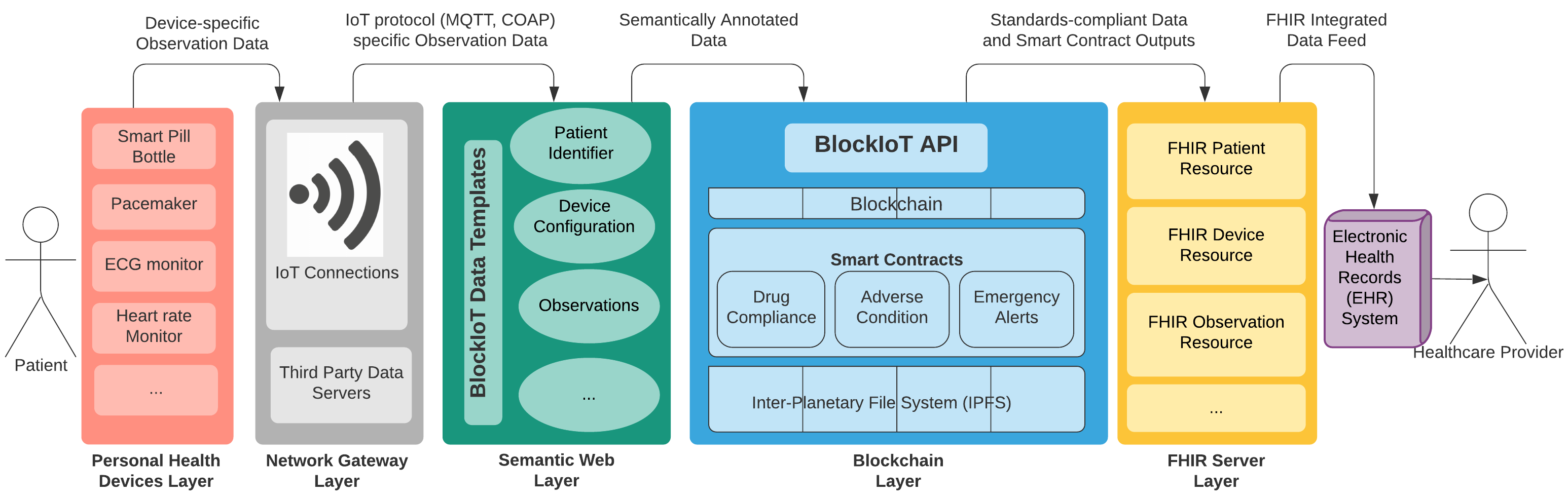}
\caption{\label{fig:data-integration} High-level Overview of the Data Streams Integrated into the BlockIoT System}
\end{figure}

\noindent\textbf{System Architecture}
The BlockIoT system is constructed as a combination of an API and various smart contracts that connect existing EHR systems to several medical devices (Figure \ref{fig:data-integration}). It is primarily characterized as a decentralized system to 
store encrypted patient medical device data received from the API, which contains endpoints that receive data published by a patient's medical IoT device. We define a medical IoT device to be an equipment that serves to collect and transmit patient health data from the patient's location (such as the patient's home) to a specific destination (such as commercial servers of the medical device company) using IoT. For example, a medical device can be a wearable heart-rate monitor that can transmit real-time heart rate to a specific destination over WiFi.

Before describing the system's implementation, some definitions are introduced first to establish the role of each component in the system. 




\textit{BlockIoT API }
The API component is created to serve as a receiving endpoint for all incoming medical device data. Because of the immense number of medical devices found in the market today, the API was created with flexibility in mind. As a result, multiple communication protocols such as HTTP(S), MQ Telemetry Transport (MQTT)~\cite{katsikeas2017lightweight}, and Constrained Application Protocol (CoAP)~\cite{bormann2012coap} are implemented in the API to ensure maximum possible coverage of all medical devices present in the market.
    
\textit{Decentralized Storage } A decentralized network was chosen (rather than a centralized system) to be the most optimal solution to ensure security, immutability, and efficiency when storing the patient data. We implemented the storage for BlockIoT using the InterPlanetary File System (IPFS) along with the InterPlanetary Naming System (IPNS), in which each patient is connected through a peer-id that is generated based on their biometrics (first name, last name, and date of birth). IPFS  was the primary choice of the BlockIoT system due to its ease of use and its ability to make data resilient over time. However, because the IPFS hashes change as the content of the file changes, IPNS is used to ensure that the address of patient content is the same regardless of the patient data change.

\noindent\textbf{System Features} 
The system is divided into two primary sections, including a medical device uploading data to the decentralized storage system, and an EHR requesting data regarding a certain patient.

\textit{Security } Compared to more popular and wide-spread centralized systems, a blockchain network architecture is likely to be more secure due to the location of data storage. Because data is stored between trusted peers rather than in a central location, the risk of patient data being leaked in wholesale from a central location is very minimal. Patient data is stored through asymmetric key encryption, in which keys to the data belonging only to two parties, i.e., the physician's EHR for a specific patient and the BlockIoT system, prevents an unauthorized user from accessing patient information. 
At the same time, patient data is only available for a specific amount of time before data access expires. This is achieved through smart contracts, which provide an IPNS link to the physician once the EHR system makes a request, and a transaction is created on the blockchain. After a certain period that is denoted on the contract, this link will expire, and another request must be made for extended access. 

\textit{Immutability} One of the main concerning aspects of centralized servers is that it is relatively easy to mutate incoming data, which can compromise the stored medical device data. The ledger in the underlying blockchain in the BlockIoT system solves this problem, as the data added to the distributed ledger is permanently stored, and any modification of the given data is discouraged, which ensures data integrity throughout BlockIoT.

\textit{Storage} In medical devices, the amount of data collected about a patient can be generated as fast as once per second. This rate extended to a larger population of 10,000 patients can reach up to millions of different data points per minute, which is enough to overwhelm most, if not all, centralized servers over time. This issue can be alleviated through a blockchain, which is server-less and can distribute the storage capacity necessary for all of these data points throughout its users. Through this system, enormous amounts of data can be stored without significant risk of a crash. 

\noindent\textbf{Template-driven Data Harmonization} 
Templates are methods that are used to analyze and identify incoming data. Because each medical device sends values with different keys, templates are essential in determining what type of data is transmitted from the device. Secondly, based on the device's identifiers, one can verify the device based on its output itself.
Each type of device will have its own template. For example, a heart rate sensor will have a template that will contain all types of information transmitted by the heart rate monitor (such as beats per minute and SpO2). The upper limit and lower limit values are also present to recognize whether a specific incoming value is within or outside a normal limit, leading to alerting the physician if the patient's condition is life-threatening. These templates are essential in BlockIoT because not only are they used as a template for data storage, but they can also serve as labeled anonymous data that can serve as training data for AI/ML algorithms for that specific physiology. 

While it is possible to send over raw medical device data to an EHR system, the data sent to the EHR system are not useful to the physician as is. It is better to facilitate the physician by providing them with vital statistics or charts. These summarized outputs enabled through smart contracts can quickly allow a physician to review the patient's health metrics and decide further treatment as necessary.
To facilitate this process, based on templates created for each type of data, physician-accessible graphs are generated and are exported to the EHR system. AI/ML algorithms are used to further determine trend lines and pinpoint metrics of interest. These algorithms can be trained to generate greater insight with real-time data, which may be further useful for a physician. 

\noindent\textbf{Making BlockIoT FHIR-Proof}     
As interoperability is a highly desired feature in BlockIoT and because we want to reach as many medical devices as possible, we designed BlockIoT to be compliant with existing standard EHR protocols that are widely used today. By being able to replicate and accommodate to specific standards that are used by a majority of the systems that are used in the healthcare industry, such as HTTP, one can state with a certain level of confidence that the proposed system will be able to function in a real-world scenario. However, we took an additional effort to make the BlockIoT system primarily compliant with the Fast Healthcare Interoperability of Resources (FHIR) Standard~\cite{bender2013hl7}, a system of requirements and guidelines on EHR system development created by the standards organization HL7. 
%
We closely examined the HL7 FHIR documentation~\cite{bender2013hl7} and leveraged the concepts related to medical devices and observations to the BlockIoT system. 

Current data in the BlockIoT system is stored in the IPFS system in a folder with various JSON files (depending on the number of medical devices that a patient has) containing patient data ordered by a timestamp. In this case, the file itself will serve as a patient resource designed to be interchangeably used between the BlockIoT system and the EHR system. Each resource contains a logical ID assigned upon the creation of the file.
The resource is created as a JSON representation with various properties that describe the patient's biometrics, the template used to identify the data, and the medical device data the patient has. A resource may also be an image that contains an interpretation of the data, such as a graph or specific data points a physician can examine before seeing the patient. The image resource is only created when an HTTP request is created to ensure that data is the most recent and relevant as possible.
    
Communication with a standardized EHR system is done through a RESTful API, which consists of various verbs used to transmit and receive resources (as defined above) from one system to another. Following this API protocol will ensure that transmittance is reliable and the incoming data is usable by a standard EHR system. The current system can conform to existing API requests properly as the routes present in the current system are identical to the routes described in the FHIR documentation. 
As a result, because the architecture of the resources and the syntax of the API conforms to the FHIR documentation, one can ensure with some confidence that the BlockIoT system will successfully communicate with an existing FHIR-based EHR system. 

\noindent\textbf{Blockchain and Smart Contracts} 
To implement smart contracts, and store transactions in a decentralized fashion, a blockchain system is implemented. In such an architecture, the nodes are the servers run by major third parties participating in this system, such as an EHR system at a hospital, manufacturer, and researchers who will employ smart contracts to send or receive patient data. Each type of node has a different level or type of access to data. EHR systems can only request data about a specific patient, for example. A smart contract will then be placed that allows physician access to the patient's data representation(in the form of an IPNS link) for a 24 hour time period. After this time access is complete, the link will become non-existent, and another request would have to be made. 

Smart contracts will then be used to analyze patient data and compare data with specific templates. For example, the ``adherence" smart contract will be programmed to examine data and check for timestamp discrepancies, and therefore determine whether the patient has missed a dose of medication. Such smart contracts can also be used for more complex data types, such as EKG monitoring, where external or industry-standard algorithms can be used to examine EKG waves, and detect whether a heart arrhythmia is present. 
Through the use of smart contracts, passive data transmission can lead to more proactive interventions. Based on physician and patient preferences (which would be present in the patient's configuration file), the smart contract can then send alerts to the relevant party if the algorithm detects an abnormality. For example, a physician can choose to ignore sharp increases in a patient's blood sugar level if the patient had just had a meal. If this option is selected by the physician, the smart contract would first detect an increase in sugar levels, text the patient if he/she had just had a meal, and the patient's response to determine whether an alert is necessary to send to the physician. Both physicians and patients can also choose when and how to receive alerts as well. 

\section*{BlockIoT In Action}

The BlockIoT model of data collection was evaluated with synthetic patient health data to test for practicality and usability. The data used for the system simulation was generated through Synthea~\cite{walonoski2018synthea}, which is a patient population generator commonly used to produce synthetic data based on distributions learned from real patient data. We tested BlockIoT against 1000 generated patients, which is typical of the number of patients associated with a physician's practice. These patient records contained different types of data, such as systolic/diastolic blood pressure, heart rate, body height/weight to represent basic vital signs, and more sophisticated lab results such as hemoglobin levels leukocyte/erythrocyte counts.
    
First, the synthetically generated patients were used to determine whether uploading and downloading from IPFS is reliable, and a decentralized system can be used as a method to store medical device data. Out of the 1000 patients simulated, all 1000 patients were successfully securely uploaded to IPFS, and medical data was reliably stored. There was no missing data present, and no packet losses occurred when uploading data. 
Next, the BlockIoT model was evaluated with multiple types of medical data, as described above. Templates for each type of data were created, and the system was checked to see if any errors with data handling occurred. Out of the 8 different types of data tested (blood pressure, heart rate, blood oxygen, sugar levels, EKG, compliance, spirometry, and cell counts), the system handled all 8 types of data appropriately, in which, the data points that were not within normal limits were recognized and represented adequately in the physician user interface. Secondly, each data point was properly stored by timestamp, and the system also captured trends over time. 
Finally, the BlockIoT system was evaluated for the ability to receive a significant amount of incoming data at once. Because medical devices tend to send information extremely rapidly (in terms of seconds), the system must handle incoming requests quickly and efficiently to ensure that all data can be uploaded in real-time. 10,000 requests were to the system at a rate of  0.5 seconds for each request. The system was able to process each request within 5-6 seconds and handle all the incoming requests without delay. 

\section*{Participatory Design Survey}

Fourteen healthcare providers(thirteen physicians and one nurse practitioner) were invited to review the BlockIoT System in order to understand the need and impact of medical devices in a clinical setting, and determine whether there is potential in patient outcome improvement that can be attributed to integrating a system such as BlockIoT in their clinical practice. While we acknowledge that the sample size is not large enough to make a statistically significant conclusion about BlockIoT's effectiveness, this evaluation aims to determine whether such a system has the potential to be feasible and useful in a participatory design setting for clinical data collection and usage scenarios.
The providers were recruited through professional and personal relations of the researchers. However, a particular focus was put on inviting physicians from different specialties. As a result, the total sample of physicians consisted of a cardiologist, a gastroenterologist, a dermatologist, a geriatrician, an emergency physician, a surgeon, an anesthesiologist, a pulmonologist/allergy/sleep medicine physician, internal medicine specialist, and multiple primary care physicians. During the recruiting process, no prior information about the project was provided to the participants, including the concept, demonstration, or questions regarding BlockIoT. We followed this procedure to ensure authenticity during the interviews. 

The recruited providers were given an option to either complete a form, or attend a virtual interview if they had time. The providers were not offered any remuneration for their answers. We conducted interviews through WebEx, and with permission from the providers, we recorded the sessions and later transcribed the discussions for analysis. For the forms, a Google Form (available at \url{http://bit.ly/blockiot-evaluation}) was used with a recorded demonstration of BlockIoT, descriptions and questions.

During the survey (for both survey types), each provider was shown a presentation describing the BlockIoT system. Afterward, the physician was shown a mockup of a sample EHR system with BlockIoT implemented. The mockup was designed to seem like an EHR system, which the physician would be most likely familiar with and therefore will easily visualize the integration.
Each provider was also provided with a HPI of a sample patient. Table \ref{hpi} summarizes the patient's health status.

\begin{table}[]
\begin{tabular}{|l|l|l|l|}
\hline
\textbf{Patient Basics} & \textbf{Diagnoses} & \textbf{Devices}                                                                         & \textbf{Current Medication} \\ \hline
Wendy Barnes         & Diabetes                & Smart Blood Sugar Monitor & Metformin  \\ \cline{2-4} 
(Age: 52; 80 bpm/98) & Obstructive Sleep Apnea & CPAP Machine              & N/A        \\ \cline{2-4} 
                     & Hypertension            & Smart Blood Pressure Cuff & Lisinopril \\ \cline{2-4} 
              & Asthma             & \begin{tabular}[c]{@{}l@{}}Smart compliance tracker\\ (for each medication)\end{tabular} & Symbicort, Albuterol        \\ \cline{2-4} 
                     & Obesity                 & Smart weight tracker      &            \\ \hline
\end{tabular}
\caption{Patient HPI Summary}
\label{hpi}
\end{table}

\begin{figure}[!htbp]
\centering
\includegraphics[width=\textwidth]{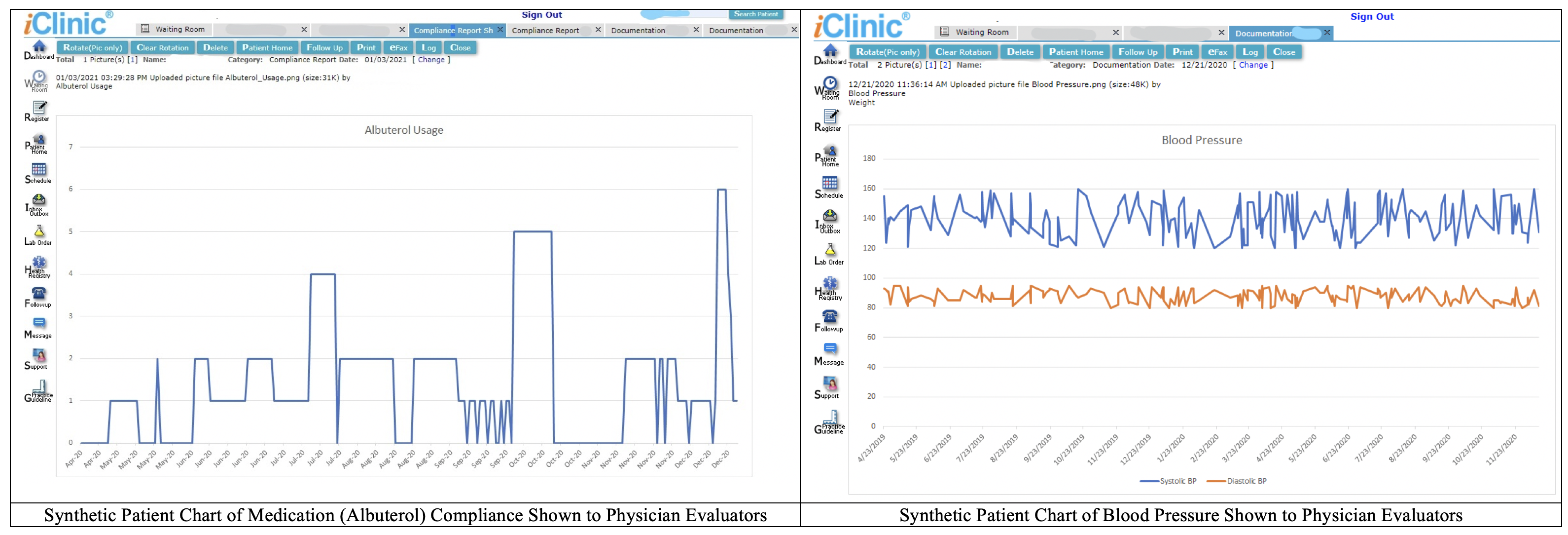}
\caption{Synthetic Patient Charts from BlockIoT Shown to Evaluators}
\label{fig:bp}
\end{figure}

The sample patient described in the HPI was designed to have multiple complications, such as obstructive sleep apnea, asthma, high blood pressure, obesity, and diabetes. This detailed patient profile was compiled to accommodate each specialist who, in this case, will be treating that aspect of the patient (for example, a cardiologist will treat the patient's blood pressure related issues). Secondly, different diseases tend to be associated with different types of medical devices. For example, a smart blood sugar monitor's purpose and method of function are different from a smart blood pressure cuff. As a result, by understanding these different devices' uses by different specialists, one can tailor the data representation through the specialty.
Furthermore, the HPI was generated to be very realistic to a typical patient these providers would see in their clinical practice. While the patient is complicated, each diagnosis is associated with the other potentially co-morbid conditions. For example, increased risk of obesity is associated with high blood pressure, diabetes, asthma, and obstructive sleep apnea, which is further corroborated because typically, patients who have diabetes are more likely to develop cardiovascular disease. As a result, the other four diagnoses were also included in the HPI. Both correlations above were used to create the HPI of the sample patient. 
The provider was then shown 2-4 graphs representing sample medical device data (such as blood pressure readings, blood sugar levels, and compliance rates). The graphs shown to the provider depended on their specialty. For example, a pulmonologist would be shown a graph with the patient's albuterol use and medication compliance in the last 30-60 days, and their blood pressure chart (Figure \ref{fig:bp}).



During each presentation, the provider was asked what difference was present between what is shown in the HPI and what the medical device data provides based on the information presented in the charts. We asked these questions to determine whether any significant impact is visible between the additional medical device data and whether this additional data can potentially change the provider's medical decision. Secondly, the provider was also asked if the graph's representation could be improved (such as changing the time axis or changing the type of graph present). This feedback can be used in future work, where the type, amount of data, or the time at which data arrives can help drive how the medical device information is represented to the provider. Moreover, each provider was asked about their overall review of the system and potential foreseeable drawbacks. At the end of the survey, for providers who were virtually interviewed, implications of the system in insurance claims, malpractice, medical data transfer from providers to providers, and potential impacts on the clinical workflow (in terms of efficiency) were also discussed.

We analyze the response data using a qualitative approach. In general, the providers' responses were positive. All of the providers found the proposed system useful, with 87\% of providers reporting that they would be willing to use the summary views as well as the patient/physician interventions throughout the patient care (the others were more inclined to just use the summary views of the medical device data). When asked whether the data coming from the BlockIoT system can change or influence the way the provider treats a patient, all the providers responded in agreement. Out of the providers asked about the efficiency impact of BlockIoT, all the providers expected the system to streamline the workflow in the long run.
%
%
Most of the providers stated that there are different levels of trust in the patient supplied information, and they use a variety of techniques to ascertain the information supplied to them by the patients during the visit. All the providers agreed that having longitudinal physiological data is better for the practice and makes them more efficient in the diagnosis process leading to better patient outcomes. However, at the moment, access to means for obtaining such data is minimal or non-existent. Physician 4 mentioned that they institute a log for patients with chronic illnesses; however, they admitted the quality of the data in these logs is only ``about 50\%", which is not sufficient enough to make an accurate health suggestion. BlockIoT would be a trustworthy mechanism in which such data could be collected in a minimally intrusive way for the patients.
Physicians 12 and 13 confirmed that having access to long-term blood pressure data for example, could help make them determine if a higher reading at the clinic is due to the ``white coat hypertension'' phenomena~\cite{pickering1988common}, especially if the readings taken at home  as available through BlockIoT appear to be lower.

Confirming the main focus of BlockIoT, the need for more data-driven analyses in clinical practices was best summarized by Physician 3.
\begin{quoting}
The only way we know whether the patient has high blood pressure is whether the patient takes a one-time blood pressure reading a day at home in the morning, and that is only when we really ask to do it. Even then, they will only do it for some of the time, not all the time. Or when they come to the doctor's office every 2 to 3 months. A blood pressure reading can change every second, which means that we are missing millions of data points right now!
\end{quoting}

However, access to this device data is also variable, i.e., no standardized method exists currently. As best stated by Physician 1 when asked about retrieving data from various medical devices:
\begin{quoting}
    It totally depends on the device; blood sugar monitors give you readouts, pacemakers need to be interrogated for data, blood pressure cuffs need to be brought to the office.
\end{quoting}

Physician 2 also raised their concerns about the lack of integration of many of the devices patients use.
\begin{quoting}
    For devices like the Apple Watch and Fitbit, there is no official way to get those reports. It's essentially the patient showing you, like `this is what my Apple Watch said' or `this is how much sleep I've been getting.' For FDA-approved devices by Medtronic, Boston Scientific, those types of companies, the office will receive data from the companies, and it's sort of like an additional provider level. But for other consumer devices, there is no specific support.
\end{quoting}

To our question on AI-driven patient-centric notifications endorsed by the provider, all of our evaluators expressed enthusiasm. In particular, physician 2 said;
\begin{quoting}
    It makes the patient-physician interaction more dynamic and more trustworthy, and the patient's getting more involved with their care, and they're more likely to be compliant with diet or lifestyle change or medication compliance if they know that they're doctors following up with them.
\end{quoting}

Physician 4 also mentioned that there would be some positive behavior change aspect when the patient knows that their data is being collected. 
\begin{quoting}
    It has to be easy for the patient because the patient knows `hey, somebody's watching me. I can't cheat. I can't do it.' This can make a huge difference.
\end{quoting}

A specific interest the providers had was to not simply look at the analyses provided through the AI algorithms in their workflows but to also pinpoint certain aspects of the data between different time-frames, to look at anomalies presented in the data analysis closely, and interrogate the data. To this end, physician 2 said;
\begin{quoting}
    We talked about having the ability to zoom in on certain times of day for certain parameters. I think that would be helpful, but it's all going to come down to how easily I can access this information.
\end{quoting}

However, some of the providers touched on the regulatory aspects of such a data-sharing ecosystem. In particular, physician 3 said;
\begin{quoting}
    You have to make sure that when there's a ton of data, there is always the patient's approval or consent that is needed so that there is not an automatic transfer from my office to the hospital. You really have to have consent. In some countries, you will find it a bit tedious in terms of getting consent, while in other health systems, it is going to be easy. 
\end{quoting}

The consensus on using the data supplied by BlockIoT for insurance claims, in authorizing tests, and in malpractice lawsuits had mixed reactions from the providers. Most of their concerns were centered around the fact that the current medical landscape is extremely complex, and any technological changes will have to be supplemented by regulatory interventions, and those changes will take some time to take effect.

\section*{Conclusion}
Within the last decade, the rise of EHRs and medical devices has revolutionized the healthcare industry by altering how to capture and manage health information. Based on our participatory design survey, however, it appears that the information from an EHR is not nearly enough to result in an optimal clinical decision. For instance, there is no concrete method of determining or verifying how a patient's health has evolved within a few months before a medical visit, as pointed out by most of our evaluators. Instead, most physicians determine diagnosis or set up treatment plans based on a snapshot of symptomatic information captured during the patient encounter or the patient's subjective description, which presents a need to access more granular and real-time patient health information. As stated by all our evaluators whose specialty involves the usage of medical devices, data from these devices can be instrumental in determining compliance, detecting problems with patients whose disease has an asymptomatic nature (such as cardiac-related problems), or figuring out the change in patients' health since the last visit. 
In traditional EHR systems, the accessibility to patients' daily health data at a granular level is still extremely limited due to centralized and proprietary medical devices that cannot share patient data with third parties. In our survey, more than half of the evaluated physicians are unable to retrieve medical device data in their current clinical practice. Among those with limited access, they have to either ask patients to physically bring in the data or obtain the data via specific ties with the device companies. 

To address these challenges, we propose and implement a universal, secure, decentralized bridge between patient-facing medical devices and physician-facing EHR systems, called \emph{BlockIoT}. It is designed to provide real-time medical device data to physicians, who are usually constrained by their busy schedules, in an easy-to-understand manner so that they can use this data to provide higher-quality and more-efficient care. As per the evaluation, BlockIoT has shown potential in improving quality of care, patient-physician interaction, office visit efficiency, and overall more informed physicians' decision-making. BlockIoT may also help physicians rule out certain diseases by offering additional data that underlie patients' health to reduce time and costs that would have been spent on determining the diagnosis. 
Moreover, BlockIoT may also contribute to other aspects, such as justifying reimbursements or reducing malpractice suits due to its ability to provide objective and concrete patient data.  
%
%
A key advantage of BlockIoT is offering physicians an integrated look into data streams---that capture information from closely monitoring patient daily life and are traditionally unavailable to them---in a relatively easy and reliable way. Data from such an integrated system are more likely to form a complete and accurate health profile of patients, which could help improve the ML algorithm that is currently mostly based on biased, truncated or even distorted data. The data supplied to the EHR systems by BlockIoT contains more specific health information that could come from a variety of patients and can potentially be incorporated into a training algorithm without excessive efforts via the proposed system. The findings may assist physicians in making more viable treatment plans for the underrepresented population and even help reduce health disparity between groups.
We believe that the proposed system, combined with the emerging technology like AI/ML, could have more far-reaching impacts. As pointed out by our physician evaluators, closely tracking patients' health evolvement and offering timely interventions and reminders could be extremely valuable but is currently very challenging given the already-heavy workload. An automatic alert system facilitated by BlockIoT and supported by AI/ML technology could be transformative, let alone the potential feature of constantly-updated parameters thanks to the improving predictability of the training algorithm after incorporating the data from medical device.

The current version of BlockIoT, however, is far from perfect. Based on the providers' interviews, the user interface and accessibility were the primary focus when asked about the potential drawbacks of the system. The consensus was that the system has the potential to be very useful to them in their clinical practice, as long as the data is easy to use and versatile enough to fit the specialist's needs. A consent system is also necessary to ensure that patients can consent to what data should be shared with the physician and how long access should be present. After the system undergoes all possible upgrades, the system's value needs to be further justified using a larger sample with sufficient statistical power. Also, there remain significant challenges to persuade stakeholders to participate in the system due to varying incentives. 
%
%
While the proposed system is not a panacea for the issues plaguing the healthcare industry, we believe that our system establishes a building block for future research and enables a step forward in addressing all the challenges faced by healthcare providers in their clinical practice, ultimately leading to better patient outcomes.

\textbf{Acknowledgements}
This work is partially supported by IBM Research AI through the AI Horizons Network and the Flash Grant from the School of Humanities, Arts, and Social Sciences (HASS) at Rensselaer Polytechnic Institute.
Special thanks to the following physicians for providing their expert opinion: Dr. Anant Agarwalla, Dr. Neha Agarwalla, Dr. Vipin Agarwalla, Dr. Preeshini Fernando, Dr. Dara Fuentes, Dr. Danekka Loganathan, Dr. Maya Matsumoto-Bessam, Dr. Adriana Miranda, Dr. Arbol Ortiz, Dr. Bharati Reddy, Dr. Benjamín Reyna Sánchez, Dr. Mayank Shukla,  Dr. Ransirini Wijeratne-Fernando and nurse practitioner Cynthia Jean-Baptiste. 

\bibliographystyle{unsrt}
\bibliography{references}

\end{document}